# Landscape Geometry-based Percolation of Traffic in Several Populous Cities around the World

Fisca Dian Utami, Dui Yanto Rahman, Desyana Olenka Margaretta,
Euis Sustini, and MikrajuddinAbdullah(a)

Department of Physics
Bandung Institute of Technology
JalanGanesa 10 Bandung 40132, Indonesia
(a)Email: mikrajuddin@gmail.com

**Abstract**

We described the average traffic congestion in several populous cities around the world from a new concept, namely landscape percolation. The ratio of the residential area size to road width is a fundamental parameter that controls the traffic congestion. We have compared the model with data extracted from several populous cities around the world (directly from Google Earth images) and demonstrated very consistent results. The criterion for a city landscape that makes a city is considered as congested or less congested has been identified. The model also explains remarkably well the consistency of the measured data with various reports on congestion levels (such as the recognized Tomtom congestion level or Numbeo traffic index) of some populous cities around the world. These findings may help in designing new cities or redesigning the infrastructure of congested cities, for example for deciding what is the maximum size of the residential area and how width the roads are. This work also shows the similarity of the problem in conducting composite (electric current flow), brine transport between icebergs (fluid flow), and traffic (vehicle flow).

# 1. Introduction

Traffic jams have become a critical problem faced by populous cities around the world. An increase in the number of vehicles that by far exceeds the expansion of roads has resulted in heavy congestion on many road lines, causing inconvenience in various aspects, such as energy inefficiency, air and sound pollution, etc. (Abdullah and Khairurrijal, 2007). Traffic transition from a free flow to a seriously congested state occurs on a daily basis in populous cities, and deteriorating the system's efficiency (Wang et al., 2015).

To understand vehicular motion comprehensively, real-time data measured in the space and time domain must be employed and spatio-temporal analysis is extremely important (Kerner, 2017). Several models have been proposed to explain the traffic congestion and most of them described how the density of vehicles change spatio-temporally (Wang et al, 2015; Kerner, 2017). Using macroscopic approach, Velasco and Marques (2005) applied a Navier-stokes equation to analyze the characteristics of traffic flow. Sumalee et al. (2011) developed the macroscopic Stochastic Cell Transmission Model (SCTM) to explain the stochastic nature of traffic flows. While for microscopic models, some authors such as Fan et al. (2018) applied a continuum traffic model integrated with an optimal velocity difference to determine the stability of the traffic flow; Yang and Monterola (2015) studied the optimal velocity function for designing the autonomous driverless vehicle system.

Wang et al. developed an agent-based model to simulate traffic using a percolation approach by including two main parameters: vehicle volume and path selection (2015). Skinner randomly designed less congestion and congestion networks in a simple lattice model and found that suboptimality degree, as an effect of the congestion, peaks at percolation threshold (2015). To date, percolation theory has been studied considerably to estimate traffic failures, obtain an efficient network and design robust networks (Wang et al, 2015; Skinner, 2015).

Now, we consider the traffic flow from different point of view. Based on how often the congestion occurs and how long extra time must be paid by the drivers to pass the roads in a city, several traffic indexes have been proposed (i.e. Tomtom congestion level (2018) and Numbeo traffic index (2018)). The index values are obtained by calculation of several factors related to traffic during a long period, and commonly one year. For example, the Tomtom congestion level calculated the average extra time spent by the driver to pass the road in a specified city during one year relative to the driving time at uncongested condition.

Therefore, in this index, the instantaneous spatio-temporal behaviors of the traffic are not considered separately. Based on this average data, the cities are given the congestion level. Since the data were collected in a very long period, we do believe that the results will be strongly affected by the geometry of the city landscape such as the width of the road and the size and shape of residential area surrounded by the roads. Therefore, it should be exist a strong correlation between the city landscape with the congestion level of the city.

To best of our knowledge, there is no report relating the city landscape with the congestion level of the city. Therefore, at the present work, we propose for the first time how the geometry of the city landscape affects its traffic behavior. We consider the city as a traffic ensemble, the concept in statistical mechanics, in which the spatio-temporal behavior of the particles/molecules is omitted. We intend to obtain general criteria for classifying cities as congested or less congested. For this reason, the model's predictions were compared to annual average traffic data, such as data from the Tomtom congestion level (2018) and the Numbeo traffic index (2018). The model, therefore, can be used in designing new cities without congestion or redesigning the infrastructure of congested cities.

We propose an approach inspired by observing the similarity between microstructure images of compacted composites of small metal particles and large polymer particles, as discussed by Malliaris and Turner (1971) and the landscapes of cities based on Google Earth images (2018). Their paper seems to be old when applied to explain the conductivity development in composites, but it is attractive when adopted to explain other phenomena like the traffic percolation as described in this work. In our model, the metallic particle is identical to vehicles on the road, and the formation of large clusters of particles represents the congestion state. Interestingly, this model has also been successfully adopted to explain the brine transport in columnar sea ice in the East Antarctic regions and the Weddell Sea (Golden et al., 1989). The flow of brine along the columns between icebergs is similar to the flow of vehicles between residential area, so that the Malliaris and Turner model is reasonably adopted to explain the traffic flow.

2. **Model**

Here we used circles instead of spheres since we are dealing with two-dimensional situation. The large cycles correspond to residential area. However, it differs from the Malliaris and Turner model (1971), instead of using small circles to represent the metallic particles, we use small squares surrounding the large circles based on the fact that the road shape is like a belt and it can be considered to be arranged by small squares. Suppose there are $N_s$ small squares with side length $D_s$ and $N_r$ larger circles with diameter $D_r$. The small squares are distributed along the circumferences of the larger circles and only develop one layer. After compacting, the shapes of the large circles deform and the small squares remain in contact along the circumferences of the large deformed circles. . We assume that the small squares do not experience any deformation. The large deformed circles represent the shape of residential areas and the chain of small squares represents the road between the residential areas. The side of the small squares represents the width of the road.

In the Google Earth images, the residential area shape is not circular. Therefore, here we define the effective diameter of a residential area $d_r = \sqrt{4A_r/\pi}$, with $A_r$ is the area of the residential area. We define the average diameter of the residential areas as $D_r = (1/n)\sum_{i=1}^{n} d_{r,i}$ where $d_{r,i}$ is the diameter of the $i$-th residential area and $n$ is the number of regions taken into account for calculation. If the width of road surrounding the $i$-th residential area is $d_{s,i}$, the average size of the small square is selected to be $D_s = (1/n)\sum_{i=1}^{n} d_{s,i}$. The circumference of a residential area is $\pi D_r$, or the total circumference of all residential areas is $N_r(\pi D_r)$.

Here, we do not consider the direction of vehicle flow. We also take into account only the main roads, clearly observed from the Google Earth images, as indicated by a relatively wide. The small squares developing a road have originated from two contacted large circles, so that each circle contributes a half of the squares. The process is similar to closing a zipper, where the zipper beads that originated from two sides filled completely the space when the zipper is closed. If the total number of vehicles at time $t$ is $N_s(t)$, the total length of the space claimed by the vehicles is $N_s(t)D_s$. The instantaneous density of vehicles on the road is given by

$$f(t) = \frac{N_s(t)D_s/2}{N_r(\pi D_r)} \tag{1}$$

The factor of half was introduced to consider that the vehicles on the street were originated from two contacting residential areas, and similar consideration has been applied by Malliaris and Turner (1971).

The occurrence of congestion does not mean all vehicles occupy all spaces along the road and leave no space. But, the congestion occurs when sufficiently large clusters are developed although other road segments remain empty. The uncongested state occurs when the vehicle density satisfies $0 \leq f(t) \leq f_c \leq 1$, with $f_c$ is the critical density for jamming, above which the congestion occurs (Greenshields, 1935; Helbing and Tilch, 1998). The $f_c$ might be related to the percolation threshold, $\rho_c \approx 0.6-0.7$, in cluster percolation model as simulated by Wang et al. (2015) and Wu et al. (2017). Lindley (1987) developed an index based on peak hour traffic volume of urban highways. The index is calculated by comparing volume to capacity $(V/C)$, and roads with $V/C$ value higher than $0.77$, are regarded as congested and this value can also be considered as $f_c$ in our case. If $N_{s0}$ is the minimum number of vehicles to generate a congestion then $N_{s0} D_s / 2 f_c = N_r (\pi D_r)$, resulting

$$\frac{N_r}{N_{s0}} = \frac{1}{2\pi f_c} \frac{D_s}{D_r} \tag{2}$$

Now let us write $N_s(t) = N_{s0} \psi(t)$. The range of $\psi(t)$ can be determined as follows. The minimum number of vehicles is zero, corresponds to $\psi(t) = 0$. The maximum $\psi(t)$ occurs when all spaces along the road are occupied by vehicles (no space left). In this case, $N_{s,\max} D_s / 2 = N_r (\pi D_r)$ or $N_{s0} \psi_{\max} D_s / 2 = N_r (\pi D_r)$, resulting

$$\psi_{\max} = \frac{2\pi N_r D_r}{N_{s0} D_s} = \frac{1}{f_c}, \tag{3}$$

The $\psi(t)$ is specific for each country and generally depends on time. At peak time, $\psi(t) \to 1$ (congestion) and in quiet road conditions, $\psi(t) \to 0$.

The fraction area occupied by vehicles relative to the total area of a city (all residential areas and all roads) becomes

$$\rho(t) = \frac{N_s(t) D_s^2}{N_{s0} D_s^2 / f_c + N_r (\pi D_r^2)/4} = \frac{\psi(t)}{1 + \frac{\pi f_c}{4} \frac{N_r}{N_{s0}} \frac{D_r^2}{D_s^2}} \tag{4}$$

Substituting Eq. (2) into Eq. (4) we obtain

$$\rho(t) = \frac{\psi(t) f_c}{1 + \frac{1}{8} \frac{D_r}{D_s}} \tag{5}$$

The average of area fraction is $\langle \rho(t) \rangle = f_c \langle \psi(t) \rangle (1 + D_r/8D_s)^{-1}$ with $\langle \rho(t) \rangle = (1/T)\int_0^T \rho(t)dt$, $\langle \psi(t) \rangle = (1/T)\int_0^T \psi(t)dt$, and $T$ = one year. The congestion occurs when $\langle \psi(t) \rangle \geq 1$ or $\langle \rho(t) \rangle > f_c (1 + D_r/8D_s)^{-1}$. Therefore, we define the percolation threshold for congestion (critical density) as

$$\rho_c = \frac{f_c}{1 + \frac{1}{8}\frac{D_r}{D_s}} \qquad (6)$$

We must pay attention to the difference between the fraction area occupied by vehicles relative to the total area of road, $f$, and the fraction area occupied by vehicles relative to the total area of a city, $\rho$, where in all cases, $\rho < f$. Equation (6) states that the percolation threshold for congestion strongly depends both on $f_c$ and on the ratio of the residential area size to the road size. The cities having large $D_r/D_s$ easily generate the congestion since the critical fraction for percolation is low.

Indeed, cities having the same $D_r$ might have different characteristics. For example, the population densities of cities are different. By assuming the vehicles moving on the road have originated from the corresponding residential areas, different cities will produce different congestion situations if they have different resident density. Sometimes a family (a house) has more than one vehicles and the other family does not have a vehicle. For example, in 2016, the average household in Anaheim, California has 2.05 vehicles, while in Jersey City, New Jersey, the average vehicle number belong to a household is only 0.85 (Governing, 2018). In 2016, around 28.9% of households in Baltimore, Maryland do not have vehicles, while in Pearland, Texas, only 1.4% households do not have vehicles (Governing, 2018). Some cities have a representative open space and other cities have very little open space. For example, Jakarta has only around 4.16% of open space (detik, 2018), while Vienna has green space of around 49.6% (MA23, 2017). Thus there is a parameter for distinguishing either a city has high or low car possession, either the city has large or less open space. To consider the effect of vehicle density from residential area, we proposed a new parameter. If the density of vehicle per area in the residential area $\xi$, the number of vehicles originated from the residential area satisfies $\propto \xi D_r^2 \propto (\sqrt{\xi} D_r)^2$. Therefore, to account for the effect of vehicle density from a residential area, we change Eqs. (5) and (6) to become

$$\rho(t) = \frac{\psi(t) f_c}{1 + \frac{\sqrt{\xi}}{8} \frac{D_r}{D_s}} \qquad (7)$$

and

$$\rho_c = \frac{f_c}{1 + \frac{\sqrt{\xi}}{8} \frac{D_r}{D_s}} \qquad (8)$$

respectively.

We can also consider the residential area's diameter to provide information about the length of a road. A road is longer if the residential area is smaller. The total length of the roads is manifested by the total circumferences of all residential areas. The ratio between the total circumference and the total residential area determines the specific road length, a concept that is similar to the specific surface area in particles (powder) science. The total circumference is $N_r(\pi D_r)$ and the total area is $N_r(\pi D_r^2/4)$. Therefore, the specific road length becomes $S_{rl} = N_r(\pi D_r)/[N_d(\pi D_r^2/4)] = 4/D_r$ and Eqs. (5) and (7) can also be expressed as $\rho(t) = [\psi(t) f_c]/[1 + 1/2 D_s S_{rl}]$ and $\rho(t) = [\psi(t) f_c]/[1 + \sqrt{\xi}/2 D_s S_{rl}]$, respectively.

## 3. Experiment

We employed the Google Earth images (2018) to calculate total residential area and total road width of 26 populous cities. The Google Earth and Open Source Map (OSM) provide free geographic data.

The calculation process began by determining the boundaries of the residential areas. We set the scale such that 1.3 cm in the display represented 1 km of the true size (1.3 cm = 1 km). This is the minimal scale at which main roads can be observed clearly **(Figure 2).** For calculating the area of a region, we selected an arbitrary starting point on the residential boundary followed by the '**measure distance**' instruction. We created a polygonal by clicking as many points along the circumference as possible until we were back at the starting point. Google automatically calculates the total area and length of the circumference. As comparison we repeated measurement using Gwyddion (2018) so that the resulting data conform to the actual scale.

To calculate the width of the main street, we used the same procedures by selecting a starting point on one side of the street followed by "**measure distance**" instruction and then selecting the second point on the opposite side (Google Earth view). We also magnified the image so that 2 cm display represents 20 m actual distance. We used a very large number of residential areas for calculations to obtain much better confidence (mostly 100 residence areas).

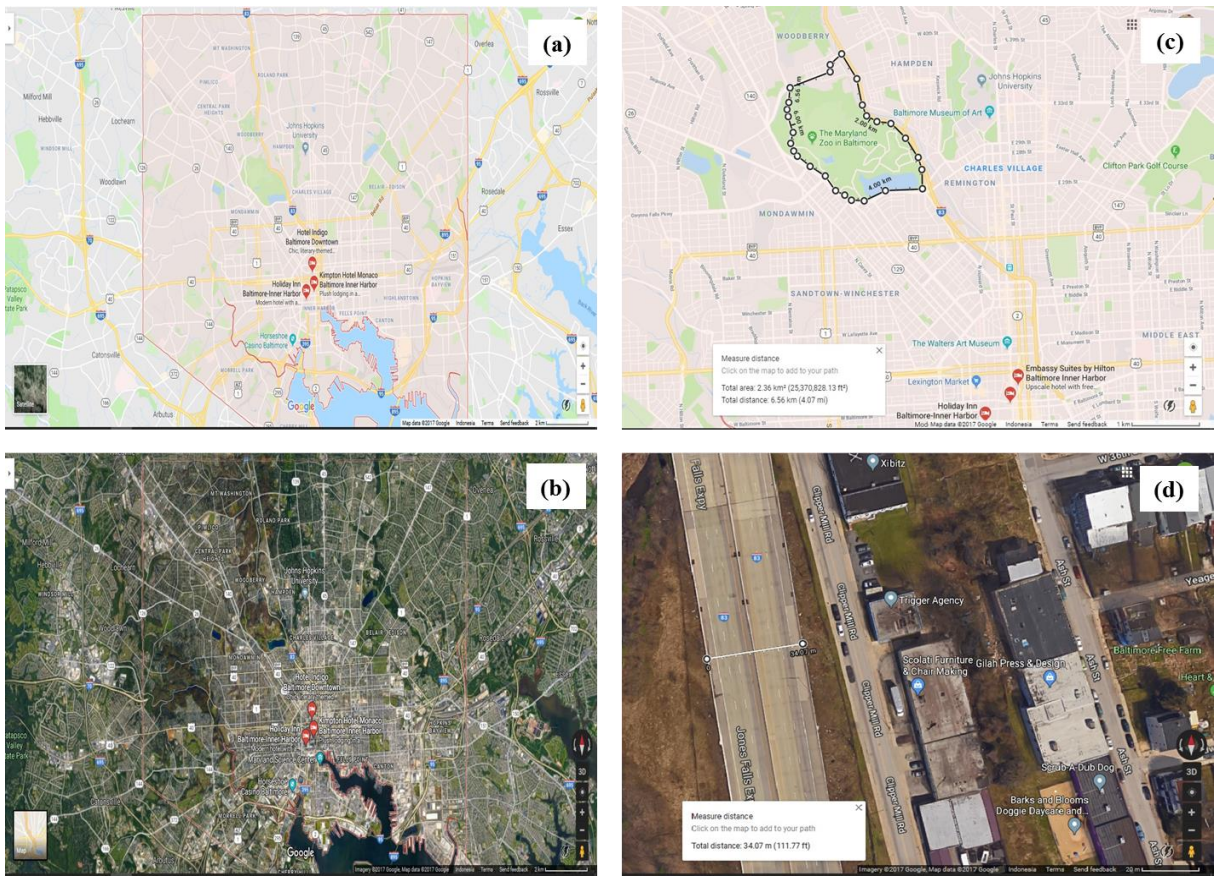

**Figure 1** (a) Baltimore in Google Map image, (b) Baltimore in Google Earth image, (c) area calculation by drawing polygons along the main road, and (d) measuring the width of a road (GoogleEarth, 2018).

## 4. Results and Discussion

**Table 1** shows the quantities belong to the examined cities. The last column shows the calculated critical fraction using Eq. (3). We were unable to obtain the *TCL* for Bandung,

Nairobi, Kolkata, Mumbai, Limmasol, Timisoara, and Reykjavik and the *NTI* for Bandung, Tainan, Changsa, Buffalo, and Grand Rapids.

**Table 1** Cities taken into account in this work and the corresponding quantities (NA = not available). The numbers in column seven have been calculated using Gwyddion (2018)

| No | City | Traffic Index (**NTI**) (Numbeo, 2018) | Congestion Level (%) (**TCL**) (Tomtom, 2018) | Number of regions taken into account, $n$ | Side of small squares (road width), $D_s$ (meter) | Diameter of large circles (region), $D_r$ (meter) | $D_r / D_s$ | $\rho_c / f_c$ |
|---|---|---|---|---|---|---|---|---|
| 1 | Jakarta | 274.39 | 58 | 98 | 18 | 1,804 (1,826) | 100 | 0.058 |
| 2 | Bandung | NA | NA | 80 | 14 | 1,548 (1,552) | 111 | 0.053 |
| 3 | Bangkok | 218.5 | 61 | 100 | 16.5 | 2,422 (2,417) | 147 | 0.040 |
| 4 | Kolkata | 283.68 | NA | 98 | 14.3 | 1,608 (1,607) | 112.3 | 0.052 |
| 5 | Moscow | 231.4 | 44 | 60 | 17 | 1,617 (1,627) | 95.7 | 0.061 |
| 6 | Mexico City | 247.72 | 66 | 100 | 17 | 1,089 (1,082) | 64.5 | 0.087 |
| 7 | Tainan | NA | 46 | 100 | 15.2 | 2,293 (2,229) | 150.6 | 0.039 |
| 8 | Rio de Janeiro | 235.57 | 47 | 100 | 21.9 | 4,406 (4,267) | 200.9 | 0.030 |

| | | | | | | | |
|---|---|---|---|---|---|---|---|
| 9 | Nairobi | 277.66 | NA | 75 | 17.9 | 2,673 (2,571) | 149.6 | 0.040 |
| 10 | Changsa | NA | 45 | 100 | 24.2 | 3,803 (2,970) | 157.2 | 0.038 |
| 11 | Mumbai | 271.95 | NA | 100 | 22.9 | 1,535 (1,679) | 66.9 | 0.084 |
| 12 | Stuttgart | 102.12 | 34 | 100 | 21 | 762 (762) | 36.3 | 0.142 |
| 13 | Berlin | 101.44 | 29 | 100 | 23 | 656 (655) | 28.5 | 0.172 |
| 14 | Vienna | 71.71 | 31 | 100 | 19 | 926 (913) | 49.4 | 0.110 |
| 15 | Paris | 161.02 | 38 | 100 | 16.5 | 372 (374) | 22.5 | 0.206 |
| 16 | Baltimore | 130.91 | 19 | 100 | 15.7 | 713 (705) | 45.5 | 0.117 |
| 17 | Buffalo | NA | 16 | 100 | 16.7 | 809 (652) | 48.3 | 0.111 |
| 18 | Grand Rapids | NA | 15 | 100 | 19.3 | 786 (811) | 40.7 | 0.129 |
| 19 | Omaha | 100.45 | 11 | 100 | 18.8 | 695 (651) | 36.9 | 0.140 |
| 20 | Dayton | 89.51 | 9 | 100 | 19.9 | 711 (670) | 35.7 | 0.144 |
| 21 | Basel | 74.00 | 27 | 50 | 18 | 597 (591) | 33.1 | 0.153 |
| 22 | Limmasol | 85.80 | NA | 50 | 18 | 613 (621) | 34.1 | 0.149 |
| 23 | Thessaloniki | 102.04 | 25 | 65 | 17.7 | 512 (504) | 29 | 0.170 |
| 24 | Timisoara | 91.51 | NA | 65 | 16.3 | 668 (688) | 40.9 | 0.129 |
| 25 | Indianapolis | 141.80 | 11 | 100 | 18.9 | 332 (290) | 17.6 | 0.246 |
| 26 | Reykjavik | 93.74 | NA | 70 | 13 | 634 (614) | 49.6 | 0.110 |

Based on congestion level (Tomtom, 2018) and media reports (Forbes, 2006; The Guardian, 2014; The Guardian, 2016; Times of India; 2018; Voanews, 2018), cities number $1-11$ can be categorized as heavy traffic (congested) cities, while cities number $12-26$ can be categorized as less congested cities. The Tomtom's congestion level (*TCL*) of the congested city is likely above 35%, while the congestion level of the less congested city is below 35%. Further, the Numbeo traffic index ($NTI$) of the congested cities is likely above 160, while the traffic index of the less congested cities is below 160. If we look at the $\rho_c/f_c$ value (last column of **Table 1**), the congested cities have an $\rho_c/f_c$ value of below 0.1, while the less congested cities have an $\rho_c/f_c$ value of above 0.1. There seems to be a strong correlation between the Tomtom congestion level (2018) and the Numbeo traffic index (2018) on the one hand and the fraction of a city's critical areas on the other hand.

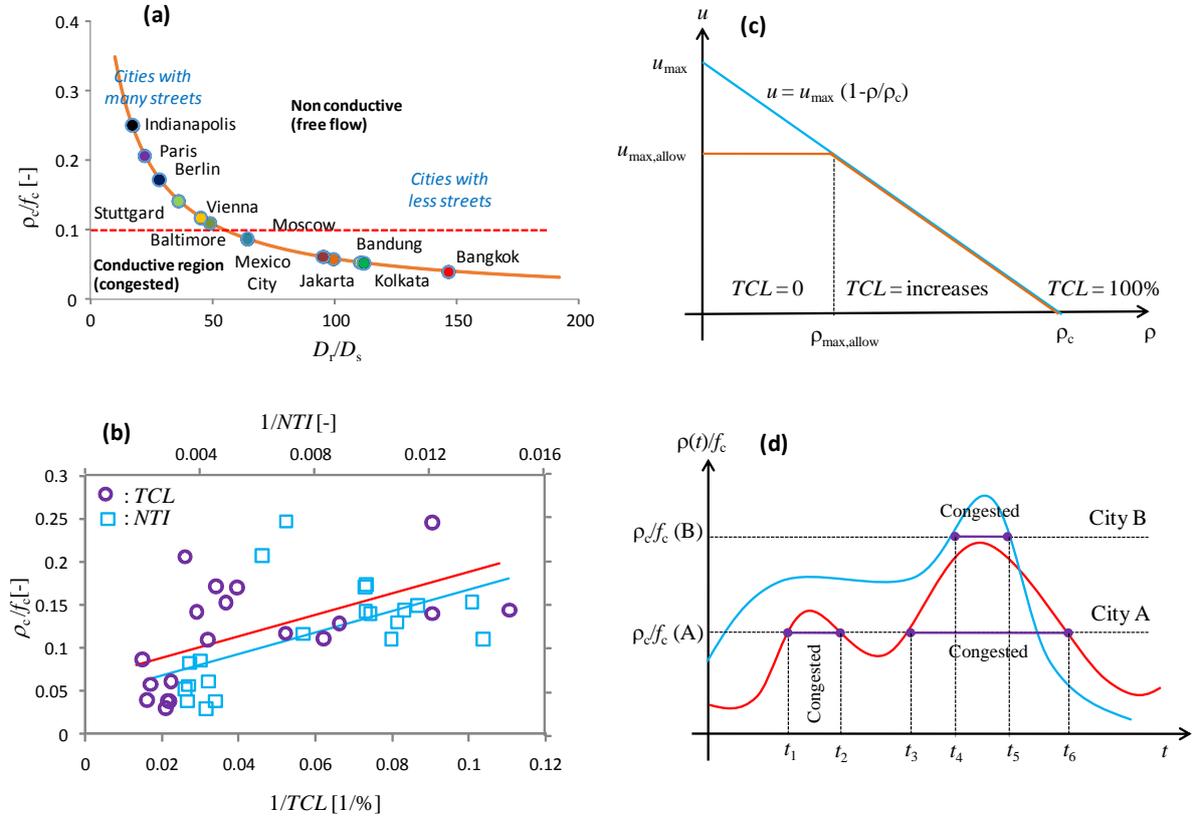

**Figure 2.** (a) Variation of the $\rho_c/f_c$ on the $D_r/D_s$ ratio and the corresponding figure for several cities. (b) The horizontal axis represents: (bottom) the inverse of Tomtom congestion level (2018), (top) the inverse of Numbeo traffic index (2018), and the vertical axis is the calculated $\rho_c/f_c$ (last column of Table 1). (c) Proposed dependence of vehicle speed on the density. The blue curve is based on Eq. (10) only. The brown curve is assumed speed based on Eqs. (10) and (11). (d) Cities having low $\rho_c/f_c$ are very easy to enter the percolation condition even at low vehicle densities while cities having high $\rho_c/f_c$ are rarely enter the percolation condition even at relatively high vehicle densities.

Variation of the $\rho_c/f_c$ on the $D_r/D_s$ ratio is shown in **Fig. 2(a),** where the $\rho_c/f_c$ decreases with increasing $D_r/D_s$ ratio. Lower $\rho_c/f_c$ represents easiness of congestion. Even with a small number of vehicles, the congested state still occurs. The lower $\rho_c/f_c$ is due to the less number of roads in a city, which means the city is dominated by residential regions. The average effective diameter of the residential areas is very large and when all vehicles originated from residential areas are using the roads, traffic stagnation occurs. In contrast, a large $\rho_c/f_c$ is obtained in cities with many roads. In these cities, the size of the residential areas is not large and the number of representative roads is very high. In **Fig. 2(b)**, we also show the position of Mexico City, Moscow, Jakarta, Bandung, Kolkata, Bangkok, Baltimore, Vienna, Stuttgart, Berlin, Paris, and Indianapolis. It is clear from the figure that the $\rho_c/f_c$ of congested cities is smaller than 0.1, while the less congested cities have a $\rho_c/f_c$ of greater than 0.1. From the examined cities, Indianapolis has the highest $\rho_c/f_c$, which means that Indianapolis is the least congested city. This is consistent with its *TCL* (2018) of only 11%. In contrast, Bangkok has a very low $\rho_c/f_c$. This indicates that Bangkok is one of the most congested cities, which is consistent with its high *TCL* (2018) of 61%.

The *TCL* (2018) measured the additional time required by the driver to pass the road compared to the time required in the uncongested condition. It depends on the vehicle velocity. The data were collected during one year and the index was obtained after averaging annual data. Suppose the measurement was conducted during $T$ (one year) or several months approaching one year. If we look at Eqs. (5) and (6), the uncongested condition does not mean that the vehicle density is zero (the velocity is maximum) since zero density means that the vehicle is absent on the road. Therefore, uncongested must mean that the vehicle can move up to permitted maximum velocity on the road (indicated by traffic sign).

For a simple model, we selected the Greenshields fuction to represent the dependence of velocity on density as (Greenshields, 1935)

$$u = u_{\max}\left(1 - \frac{\rho}{\rho_c}\right) \qquad (10)$$

The modification of this model has proposed by Helbing and Tilch (1998) by introducing a more general relationship, $u = u_{\max}\left(1 - (\rho/\rho_c)^{l-1}\right)^{1/(1-m)}$ with a best fit was obtained using $l \approx 2.8$ and $m \approx 0.8$. Suppose the permitted maximum velocity at the $u_{\max,allow}$, then the corresponding maximum density of vehicle so that the driver can drive up to this permitted velocity is $\rho_{max,allow}$ satisfies

$$u_{max,allow} = u_{max}\left(1 - \frac{\rho_{max,allow}}{\rho_c}\right) \qquad (11)$$

This means that, even the vehicle density, $\rho$, much lower that $\rho_{max,allow}$ or $0 \leq \rho \leq \rho_{max,allow}$, the speed must not exceed $u_{max,allow}$, or $0 \leq u \leq u_{max,allow}$. **Figure 2(c)** illustrates the speed of vehicle as a function of density based on Eq. (10) only (blue curve) and based on combined Eq. (10) and (11). This shape was also discussed by Jiang et al. (2001). The presence of flat section at low density was also discussed by Wang et al. (2011). If based on Eq. (10) only we have one straight line that maximum at $\rho = 0$ and zero at $\rho_c$. However, by using a combined Eq. (10) and (11) we have the speed is constant at $u_{max,allow}$ when $0 \leq \rho \leq \rho_{max,allow}$ and decreases according to Eq. (10) when $\rho > \rho_{max,allow}$.

The minimum time required by the driver to pass the road of length $L$ in the uncongested condition becomes

$$t_{uncong} = \int_0^L \frac{dx}{u_{max,allow}} = \int_0^L \frac{dx}{u_{max}(1 - \rho_{max,allow}/\rho_c)} \qquad (12)$$

However, in real situation, the velocity of the vehicle is approximated with Eq. (10) so that the time required by the vehicle to pass a road of length $L$ in real situation is

$$t_{real} = \int_0^L \frac{dx}{u} = \int_0^L \frac{dx}{u_{max}(1 - \rho/\rho_c)} \qquad (13)$$

Therefore, the additional time required by the driver becomes $\Delta t = t_{real} - t_{uncong}$. We can show easily the following result

$$\frac{\Delta t}{T} = \frac{L/T}{u_{max}(\rho_c - \rho_{max,allow})}\left\langle \frac{(\rho - \rho_{max,allow})}{(1 - \rho/\rho_c)} \right\rangle$$

$$= \frac{\kappa}{(\rho_c - \rho_{max,allow})}\left\langle \frac{(\rho - \rho_{max,allow})}{(1 - \rho/\rho_c)} \right\rangle \qquad (15)$$

with $\kappa = L/Tu_{max}$ is a constant. This equation must be proportional to the *TCL*. By selecting the appropriate parameter $\kappa$ we can express the *TCL* as

$$TCL = \frac{\kappa}{(\rho_c - \rho_{max,allow})} \left\langle \frac{(\rho - \rho_{max,allow})}{(1 - \rho/\rho_c)} \right\rangle \tag{17}$$

But we must keep in mind that when the calculation results $TCL < 0$ (caused by $\rho < \rho_{max,allow}$), the $TCL$ must be adjusted to 0 since at that situation, the density of the vehicle is less than the density for driving at the maximum allowed speed. Therefore all vehicles are assumed to drive at that maximum allowed speed. The upper bound of the vehicle density so that the $TCL$ is zero is when $\left\langle \frac{(\rho - \rho_{max,allow})}{(1 - \rho/\rho_c)} \right\rangle = 0$. In the case of city having $\rho \ll \rho_c$ we approximate

$$TCL \approx \frac{\kappa}{(\rho_c - \rho_{max,allow})} \left( \langle \rho \rangle - \rho_{max,allow} \right) \tag{18}$$

Therefore, the upper bound of the city to have zero $TCL$ is $\langle \rho \rangle = \rho_{max,allow}$. It is clear from Eq. (18) that the $TCL$ increases with the average vehicle density $\langle \rho \rangle$, approximately linearly at low vehicle density. The $TCL$ is also dependent on the traffic rule in the city since different city might have different $u_{max,allow}$ to mean different $\rho_{max,allow}$. If $u_{max,allow}$ in a city is low ($\rho_{max,allow}$ is low), the $TCL$ will be very high even the vehicle density is low but has surpassed $\rho_{max,allow}$.

Now, let us compare this prediction with data in Table 1. To investigate the relation of $TCL$ with $\rho_c$ We can rewrite Eq. (18) as

$$\rho_c / f_c = \rho_{max,allow} / f_c + \frac{\kappa(\langle \rho \rangle - \rho_{max,allow})/f_c}{TCL} \tag{19}$$

We just need to inspect how $TCL$ changes with $\rho_c$. For a rough investigation, let us assume other parameters are nearly constant so that the above equation can be approximated as $y = a + bx$ with $x = 1/TCL$, $y = \rho_c/f_c$, $a = \rho_{max,allow}/f_c$, and $b = \kappa(\langle \rho \rangle - \rho_{max,allow})/f_c$. We fit the data in Table 1 with Eq. (19). The result is shown in circle symbols in Fig. 3(b). The best fitting line (even if they seem scattered) resulting $y = 0.065 + 1.171x$. Based on Table 1, the demarcation of the congested and less congested cities is at $\rho_c/f_c \approx 0.1$. Using this data, the demarcation of the $TCL$ for congested and less congested cities is around $TCL \approx 33.5\%$, which is consistent with the $TCL$ for Stuttgart of 34 (2018) which is located at the demarcation of the congested and less congested cities.

If we assume that the Numbeo traffic index (*NTI*) also satisfies the similar relation with $x = 1/NTI$ and $y = \rho_c / f_c$ we obtain the fitting equation, $y = 0.044 + 9.142x$. Again using the demarcation of the congested and less congested cities is at $\rho_c \approx 0.1$, we obtain the demarcation of the *NTI* (2018) is $\approx 163$. This is also consistent with data in table 1 where the highest *NTI* for less congested city is Paris with *NTI* is 161. All congested cities have *NTI* of above 200. Surprisingly, the $\rho_c/f_c$ values in both fittings are nearly the same (0.065 and 0.044) although no adjustment was performed.

Mexico City, Moscow, Jakarta, Bandung, Kolkata, and Bangkok are the most congested cities, while Baltimore, Vienna, Stuttgart, Berlin, Paris, and Indianapolis are less congested cities. The first mentioned cities are very easy to enter the percolation condition (jamming) since the percolation thresholds are low while the later mentioned cities are rarely entering the percolation condition since the percolation thresholds are high. Even at low vehicle densities, the first-mentioned cities have entered the percolation condition while the later mentioned cities might stay at free flow condition at relatively high vehicle densities as illustrated in **Fig. 2(d).**

As shown in Eq. (8), a city is classified into congested or congested depends also on the density of vehicles in the city. Rural city has very low vehicle density ($\xi \ll 1$) so that even the ration of residential area size to the road width is high, the congestion rarely occurs. To the contrary, the urban city having high vehicle density ($\xi \to 1$) will easily enter the congestion condition. **Figure 3** shows the contour plot of $\rho_c/f_c$ from Eq. (3) as function of $D_r/D_s$ and $\xi$. The congested and less congested cities are separated by a curve with $\rho_c/f_c = 0.1$. The region below this curve belongs to less congested cities, while that above this curve belongs to congested cities. The curve with $\rho_c/f_c = 0.1$ satisfies the equation $\xi = 5184/(D_r/D_s)^2$. Very isolated region having very low vehicle density ($\xi \to 0$) the congestion never occurs. Theoretically, the congestion occurs when $D_r/D_s \to \infty$.

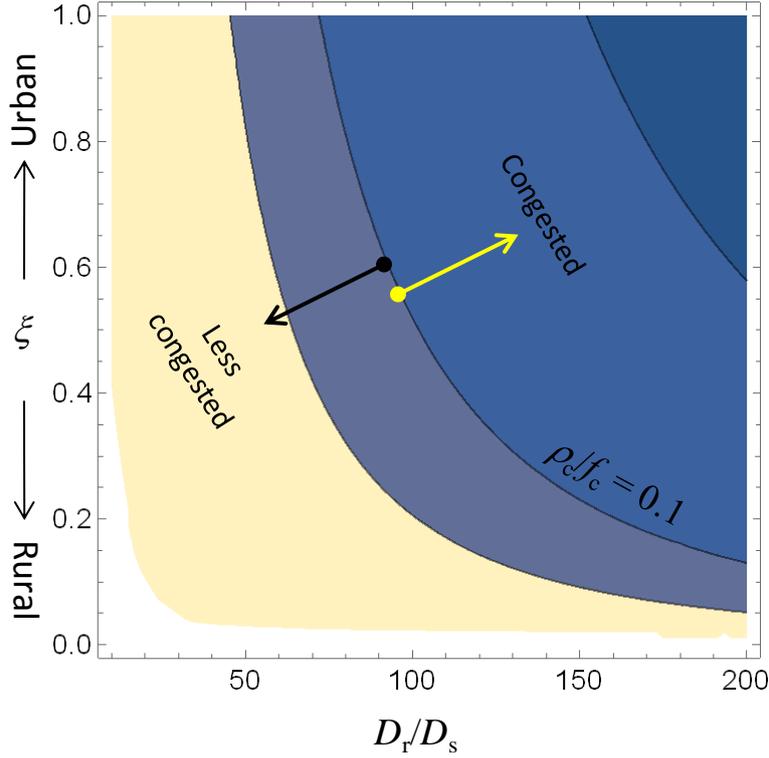

**Figure 3** Contour plot of $\rho_c/f_c$ from Eq. (3) as function of $D_r/D_s$ and $\xi$. Curve with $\rho_c/f_c = 0.1$ is a curve separating the congested (upper) and less congested (lower) cities.

It seems that the proposed model was able to identify the congestion level of several cities around the world. The main parameter is the ratio between residential area and road width. Congested cities, such as populous cities in developing countries, have a large ratio of residential area to road width, while less congested cities have a small ratio of residential area to road width.

## 5. Conclusion

The model of landscape percolation has been successfully applied to describe the traffic percolation in several populous cities around the world. We strongly identified that the congested condition of cities depends on the ratio of residential area to road width. After defining the area fraction of road, we obtained a criterion separating congested and less congested cities, i.e. the area fraction of about 0.1. Cities with an area fraction of less than 0.1 are classified as congested cities, while cities with area fraction of above 0.1 are classified as less congested cities. The conclusion is supported by several statistical data regarding the traffic conditions of cities around the world such as the Tomtom congestion level and the Numbeo traffic index.